\documentclass[aps,twocolumn,showpacs]{revtex4}
\usepackage{graphicx}
\def\bsigma{\mbox{\boldmath $\sigma$}}

\begin{document}

\title{Generation-recombination processes via acoustic phonons
in a disorded graphene}
\author{F. T. Vasko}
\email{ftvasko@yahoo.com}
\author{V. V. Mitin}
\affiliation{Department of Electrical Engineering, University at Buffalo, Buffalo, NY 1460-1920, USA}
\date{\today}

\begin{abstract}
Generation-recombination interband transitions via acoustic phonons are allowed in a disordered graphene because of violation of the energy-momentum conservation requirements. The generation-recombination processes are analyzed for the case of scattering by a short-range disorder and the deformation interaction of carriers with in-plane acoustic modes. The generation-recombination rates were calculated for the cases of intrinsic and 
heavily-doped graphene at room temperature. The transient evolution of nonequilibrium carriers is described by the exponential fit dependent on doping conditions and disorder level. The characteristic relaxation times 
are estimated to be about 150 - 400 ns for sample with the maximal sheet
resistance $\sim$5 k$\Omega$. This rate is comparable with the generation-recombination processes induced by the thermal radiation.
\end{abstract}
\maketitle

\section{Introduction}

Transport \cite{1} and optical \cite{2} properties of graphene as well as noise phenomena in this material \cite{2a} are not
completely understood for the regime of nonlinear response. The treatment of nonequilibrium carriers requires not only verification the momentum and energy relaxation processes but also understanding of the interband generation-recombination processes which determine electron and hole
concentrations far from equilibrium (similar transport conditions take place for the bulk gapless materials, see review \cite{3} and references therein). Effective interband transitions via optical phonons of energy
$\hbar\omega_0$ take place for the energy of carriers greater than $\hbar\omega_0 /2$, see Refs. 5 and 6 where the cases of optical excitation and heating by dc 
current were analyzed. At lower energies, the generation-recombination processes become ineffective because the Auger transitions are forbidden
due to the symmetry of electron-hole states \cite{6} (c.f. with \cite{7}).
Since the carrier's velocity $\upsilon\simeq 10^8$ cm/s exceeds significantly the sound velocity $s$, the interband transitions via acoustic phonons are also forbidden due to the momentum-energy conservation laws. Only slow
generation-recombination processes induced by the thermal radiation are
allowed in a perfect graphene. \cite{8} To the best of our knowledge, consideration of a disorder effect on the interband transitions via 
acoustic phonons in the low-energy region, $\varepsilon <\hbar\omega_0 
/2$, is not performed yet. Thus, the evaluation of the 
generation-recombinaton rate caused by the interaction of carriers with 
the acoustic phonon thermostat under violation of the momentum-energy laws in a disordered graphene (allowed electron-hole  transitions are depicted in 
Fig. 1) is timely now.

In this paper, the calculations are performed for the model of short-range disorder whose parameters are taken from the mobility data, \cite{1,9} for samples with the maximal resistance of 2 - 6 k$\Omega$ per square.
The probability of electron-hole transitions is expressed through the 
averaged spectral density functions and is calculated taking into account 
the contribution of interband interference. Due to slowness of the interband
transitions, the quasiequilibrium distributions of electrons and holes with
the same temperature are used for the description of a temporal evolution of nonequilibrium concentrations of carriers. The electron and hole concentrations are also connected through the electroneutrality condition with the surface charge controlled by a gate voltage.
\begin{figure}[tbp]
\begin{center}
\includegraphics[scale=0.65]{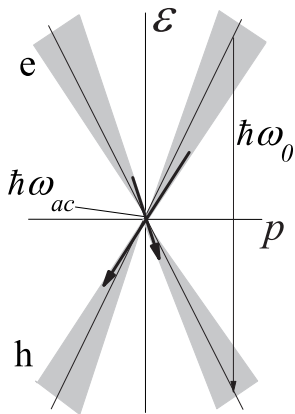}
\end{center}
\addvspace{-0.5 cm}
\caption{Interband generation-recombination transitions via acoustic 
phonons with energies $\sim\hbar\omega_{ac}$ (thick arrows) between 
broadened electron-hole (e - h, shown by grey) states in the low-energy region, $\varepsilon <\hbar\omega_0 /2$. Thin lines show the ideal 
dispersion law.}
\end{figure}

The results were obtained for the cases of intrinsic and heavily-doped
graphene at temperature $T$ and can be briefly summarized as follows. 
The concentration balance equation is written through the chemical potential normalized to $T$ and the characteristic rate, which is proportional to 
a carrier-phonon coupling and increases with temperature as $T^2$. The transient evolution of nonequilibruim population can be fitted by an exponential decay with the relaxation time 150 - 400 ns at room temperature and a typical disorder level corresponding to the maximal sheet resistance $\sim$5 k$\Omega$. This time scale appears to be comparable to the recombination rate via thermal radiation and the mechanism under consideration can be verified by temperature and temporal measurements.

The paper is organized as follows. In the next section we present the 
basic equations which describe the generation-recombination processes under
consideration. In Sec. III we evaluate the generation-recombination rates
and analyze their dependencies on temperature, disorder level, and doping conditions. The last section includes the discussion of the approximations 
used and conclusions. In Appendix we consider the generation-recombination mechanism caused by the interaction with the thermal radiation.

\section{Basic Equations}
Temporal evolution of carriers in a random potential, which are weakly interacting with the acoustic phonon modes, is described by the distribution
$f_{\alpha t}$ over the states $|\alpha )$ with energies $\varepsilon_\alpha$.
An exact with respect to a disordere effect kinetic equation takes the form 
\cite{10}
\begin{eqnarray}
\frac{\partial f_{\alpha t}}{\partial t} = \sum\limits_{\alpha '}\left[ 
W_{\alpha '\alpha}f_{\alpha 't}\left( 1-f_{\alpha t}\right)\right.
\nonumber  \\
\left. -W_{\alpha\alpha '}f_{\alpha t}\left( 1-f_{\alpha 't}\right)\right] .
\end{eqnarray}
The transition probability $W_{\alpha\alpha '}$ is written within the 
Born approximation with respect to the carrier-phonon interaction with
$q$th phonon mode of frequency $\omega_q$:
\begin{eqnarray}
W_{\alpha\alpha '}=\frac{2\pi}{\hbar}\sum\limits_q\left| (\alpha |
\hat\chi_q |\alpha ')\right|^2 ~~~~~~~~~~~~~~~~ \\
\times\left[ (N_q +1)\delta\left(\varepsilon_\alpha -\varepsilon_{\alpha '} 
-\hbar\omega_q\right) +N_q\delta\left(\varepsilon_\alpha -
\varepsilon_{\alpha '} +\hbar\omega_q\right)\right] .  \nonumber
\end{eqnarray}
Here the operator $\hat\chi_q$ determines the carrier-phonon interaction $\sum_q\left(\hat\chi_q \hat{b}_q + H.c. \right)$ where $\hat{b}_q$ is the annihilation operator of $q$th mode and $N_q$ is the Planck distribution 
of phonons at the equilibrium temperature $T$. Note, that the transition probabilities $W_{\alpha '\alpha}$ and $W_{\alpha\alpha '}$ are connected 
by $W_{\alpha '\alpha}=\exp\left[ -\left(\varepsilon_\alpha -\varepsilon_{\alpha '}\right) /T\right] W_{\alpha\alpha '}$ and
\begin{equation}
\sum\limits_{\alpha\alpha '}\left[ W_{\alpha '\alpha}f_{\alpha 't}
\left( 1-f_{\alpha t}\right) -W_{\alpha\alpha '}f_{\alpha t}\left( 
1-f_{\alpha 't}\right)\right] =0
\end{equation}
due to the particle conservation law.

The concentrations of electrons and holes, $n_t$ and $\overline{n}_t$, 
which are averaged over random disorder (such averaging is denoted as $\langle\ldots\rangle$), are given by
\begin{equation}
\left|\begin{array}{*{20}c} {n_t } \\ \overline{n}_t\end{array}\right|
=\frac{4}{L^2}\left\langle\sum\limits_\alpha\left|{\begin{array}{*{20}c}
{\theta\left(\varepsilon_\alpha\right) f_{\alpha t}} \\
{\theta\left( -\varepsilon_\alpha\right)\left( 1-f_{\alpha t}\right)} 
\end{array}} \right| \right\rangle  , 
\end{equation}
where $L^2$ is the normalization area and the step function $\theta
(\pm\varepsilon )$ appears due to the symmetry of electron-hole spectrum [see
Eq. (10) below]. Since effective intraband scattering is caused by the
phonon thermostat and carrier-carrier interaction, the quasiequilibrium distributions over the conduction and valence bands are imposed during a short-time scales and below we use
\begin{equation}
\tilde f_{\varepsilon t}  = \left\{ {\begin{array}{*{20}c}
{\left[ {\exp \left( {\frac{{\varepsilon -\mu_t^ >}}{T}} \right) + 1} \right]^{ - 1} ,} & {\varepsilon  > 0}  \\
{\left[ {\exp \left( {\frac{{\varepsilon -\mu_t^<}}{T}} \right) + 1} \right]^{ - 1} ,} & {\varepsilon  < 0}  \\
\end{array}} \right. .
\end{equation}
Due to effective energy relaxation the same temperatures are established in both bands. At the same time the electon and 
hole concentrations are determined through the different chemical potentials, 
$\mu_t^>$ and $\mu_t^<$, respectively. The chemical potentials are connected by the electroneutrality condition $n_t-\overline{n}_t=n_s$, where the surface charge $en_s$ is controlled by the gate voltage, $V_g$, according to $n_s=aV_g$ with $a\simeq 7.2\times 10^{10}$ cm$^{-2}$/V written for the SiO$_2$ substrate of thickness 0.3 $\mu$m.

The concentration of electrons is governed by the balance equation
$dn_t /dt=(dn/dt)_{ac}$ with the generation-recombination rate
\begin{eqnarray}
\left(\frac{dn}{dt}\right)_{ac}=\int\limits_0^\infty d\varepsilon 
\int\limits_{-\infty}^0 d\varepsilon 'W(\varepsilon ,\varepsilon ') ~~~~~ \\
\times \left[ {\exp \left( {\frac{\varepsilon ' -\varepsilon}{T}} \right)\left( 1 -\tilde f_{\varepsilon t}\right)\tilde f_{\varepsilon 't}  - \left( {1-\tilde f_{\varepsilon 't} } \right)\tilde f_{\varepsilon t} } \right] . \nonumber
\end{eqnarray}
We take into account that the intraband transitions (when $\varepsilon , \varepsilon '>0$) vanish in (6) and transform the transition probability as follows
\begin{equation}
W(\varepsilon ,\varepsilon ') =\frac{4}{{L^2 }}\left\langle {\sum\limits_{\alpha \alpha '} {\delta \left( {\varepsilon  -\varepsilon_\alpha  } \right)\delta \left( {\varepsilon ' -\varepsilon_{\alpha '} } \right)W_{\alpha\alpha '} } } \right\rangle .
\end{equation}
Further, we introduce the exact spectral density function
\begin{equation}
A_\varepsilon  \left( {l{\bf x},l'{\bf x}'} \right) = \sum\limits_\alpha  {\delta \left(\varepsilon -\varepsilon_\alpha \right)\Psi_{l{\bf x}}^{(
\alpha )} \Psi _{l'{\bf x}'}^{(\alpha )*} } ,
\end{equation}
which is determined through the double-row wave function $\Psi_{l{\bf x}}
^{(\alpha )}$ with $l=$1,2. The column $\Psi_{\bf x}^{(\alpha )}$ is
a solution of the eigenvalue problem $(\hat h+V_{\bf x})\Psi_{\bf x}
^{(\alpha )}=\varepsilon_\alpha\Psi_{\bf x}^{(\alpha )}$ written through
the single-particle Hamiltonian $\hat h$ and a random potential $V_{\bf x}$. Using the definition (2) one obtains the probability $W(\varepsilon ,\varepsilon ')$ as follows
\begin{eqnarray}
W(\varepsilon ,\varepsilon ')=\frac{8\pi}{\hbar L^2}\sum\limits_{\bf q}|C_q|^2\left( N_q +1\right)\delta\left(\varepsilon -\varepsilon ' -\hbar\omega_q\right) \\
\times\int d{\bf x}\int d{\bf x}'e^{i{\bf q}\cdot ({\bf x}-{\bf x}')}{\rm tr}\left\langle\hat A_{\varepsilon '} \left( {\bf x},{\bf x}' \right)\hat A_\varepsilon\left({\bf x}',{\bf x}\right)\right\rangle , \nonumber
\end{eqnarray}
where $\bf q$ is the in-plane wave vector and $|C_q|^2$ is the matrix element
of deformation interaction. \cite{9}
As a result, $(dn/dt)_{ac}$ is expressed through the two-particle correlation function. Since the main contributions to (6) appears from $\varepsilon\neq \varepsilon '$, this correlation function can be decoupled according to
$\left\langle {\hat A_{\varepsilon '} \left( {{\bf x},{\bf x}'} \right)\hat A_\varepsilon  \left( {{\bf x}',{\bf x}} \right)} \right\rangle  \approx \hat A_{\varepsilon ',\Delta {\bf x}} \hat A_{\varepsilon , - \Delta {\bf x}}$,
where $\hat A_{\varepsilon ,\Delta {\bf x}}  = \left\langle {\hat A_\varepsilon  \left( {{\bf x},{\bf x}'} \right)} \right\rangle$ is the
averaged spectral function given by 2$\times$2 matrix.

Below, we calculate the probability (9) using the model of the short-range  disorder described by the Gaussian correlator $\langle V_{\bf x}V_{\bf x'}\rangle =\bar V^2\exp [-({\bf x}-{\bf x'})^2/2l_c^2]$, where $\bar V$ is the averaged amplitude, $l_c$ is the correlation length, and the cut-off energy
$E_c =\upsilon\hbar /l_c$ exceeds the energy scale under consideration.
According to Refs. 12 and 13 the retarded Green's function in the
momentum representation takes form:
\begin{eqnarray}
\hat G_{\varepsilon ,{\bf p}}^R  = \hat P_{\bf p}^{(+)} G_{\varepsilon ,p}+ 
\hat P_{\bf p}^{(-)} G_{\varepsilon ,-p} , ~~~~~~ \\
G_{\varepsilon ,p} \approx\left[\varepsilon (1 + \Lambda_\varepsilon +ig) -\upsilon p\right]^{-1} , ~~~ \Lambda_\varepsilon =\frac{g}{\pi}
\ln \left(\frac{E_c}{|\varepsilon |} \right) \nonumber
\end{eqnarray}
where $\hat P_{\bf p}^{(\pm )}=\left[ 1\pm (\hat{\bsigma}\cdot{\bf p})/p\right] /2$ are the projection operators on the conduction ($+$) and valence 
($-$) bands, $\hat{\bsigma}$ is the isospin Pauli matrix, and 
$g=(\bar V^2 l_c/\hbar\upsilon )^2\pi/2$ is the coupling constant.
Here we restrict ourselves by the Born approximation when the self-energy
contribution $\varepsilon (\Lambda_\varepsilon +ig)$ is written through the logarithmically-divergent real correction and the damping factor.
Note, that these corrections vanish at $\varepsilon\to 0$. From
a comparision with the mobility data \cite{1,9} one obtains that the parameters $g\simeq$0.45, 0.3, and 0.15 correspond to the sheet resistances $\sim$6, $\sim$4, and $\sim$2 k$\Omega$ per square, respectively. The density of states, $\rho_\varepsilon =-4{\rm Im}\sum_{\bf p}{\rm tr} 
\hat G_{\varepsilon ,{\bf p}}^R$ is shown in Fig. 2a and $\rho_\varepsilon =\rho_{-\varepsilon}$, i. e. the electron-hole symmetry is not violate  
due to disorder. Since $\rho_\varepsilon$ increases in comparision to the ideal case, $\overline\rho_\varepsilon =2|\varepsilon |/[(\hbar\upsilon )^2
\pi ]$, the energy-dependent renormalized velocity, $\upsilon V_\varepsilon$ 
decreases up to 10\% if $g\leq$0.5 and the concentration of carriers in 
an intrinsic graphene increases up to 2 times, see Figs. 2b and 2c, respectively (here $n_T\simeq 8.1\times 10^{10}$ cm$^{-2}$ is the equilibrium
concentration at room temperature and at $g\to$0). 
\begin{figure}[tbp]
\begin{center}
\includegraphics{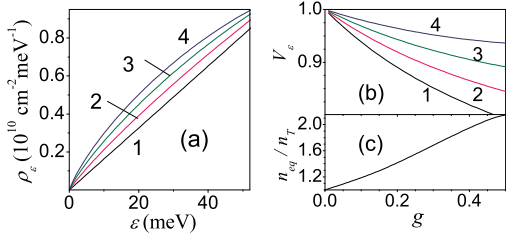}
\end{center}
\addvspace{-0.5 cm}
\caption{(Color online) (a) Density of states $\rho_\varepsilon$ versus
energy at $g=$0 (1), 0.15 (2), 0.3 (3) and 0.45 (4). (b) Ratio $V_\varepsilon =\sqrt{\overline{\rho}_\varepsilon /\rho_\varepsilon}$ versus $g$ for
$\varepsilon =$20 meV (1), 30 meV (2), 40 meV (3), and 50 meV (4). (c)
Equilibrium concentration of non-doped graphene $n_{eq}$ versus $g$ 
normalized to $n_T\simeq 0.52 (T/\hbar\upsilon )^2$.}
\end{figure}

\begin{figure}
\begin{center}
\includegraphics{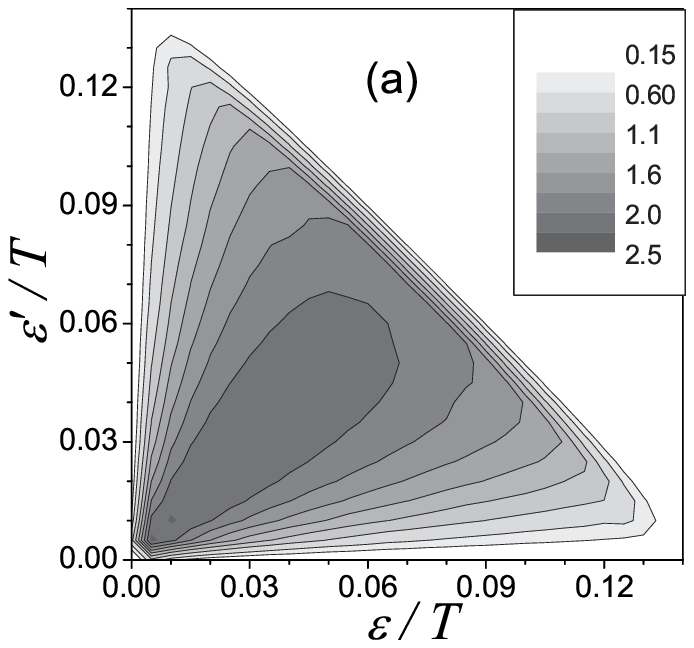}
\includegraphics{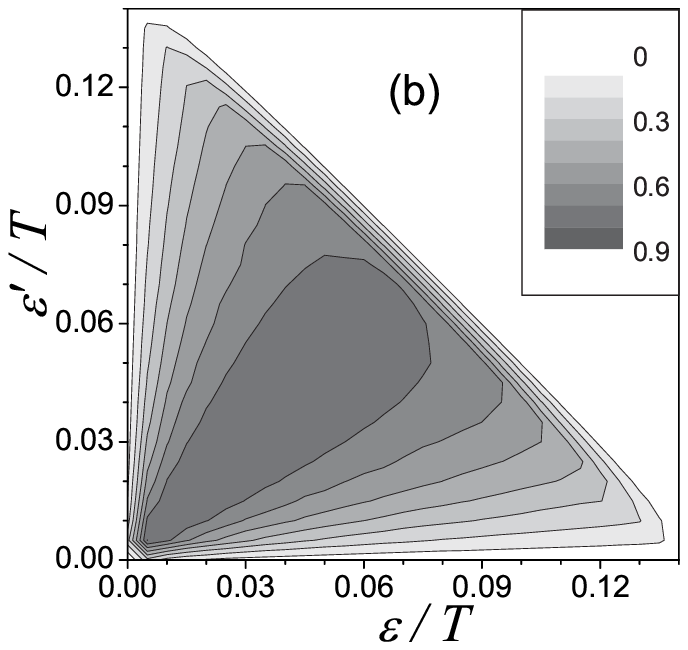}
\end{center}
\addvspace{-1 cm}
\caption{Contour plots of dimensionless kernels $w(\varepsilon /T, 
\varepsilon '/T)$ for $g=$0.4 (a) and $g=$0.2 (b). }
\end{figure}

Further, we use the standard relation $\hat A_{\varepsilon ,{\bf p}}=i
\left(\hat G_{\varepsilon ,{\bf p}}^R -\hat G_{\varepsilon ,{\bf p}}^{R~+}\right)/2\pi $ and transform the probability (9) taking into account the energy conservation law:
\begin{eqnarray}
W(\varepsilon ,\varepsilon ')\approx\left\{ |C_q |^2(N_q +1)\right\}
_{\hbar\omega_q =\varepsilon -\varepsilon '} \\
\times\frac{8\pi}{\hbar L^2}\sum\limits_{\bf pp'}\delta\left(\varepsilon -\varepsilon '-s|{\bf p}-{\bf p'}|\right){\rm tr}\left(\hat A_{\varepsilon ',
{\bf p}'}\hat A_{\varepsilon ,{\bf p}} \right) . \nonumber
\end{eqnarray} 
The trace here should be taken using ${\rm tr}\left(\hat P_{\bf p'}
^{(\pm )}\hat P_{\bf p}^{(\pm )}\right) =[1+({\bf p}\cdot {\bf p'})/pp']/2$ and ${\rm tr}\left(\hat P_{\bf p'}^{(\pm )}\hat P_{\bf p}^{(\mp )}\right) 
=[1-({\bf p}\cdot {\bf p'})/pp']/2$. This result differs from the standard 
consideration, \cite{13} because {\it interference of electron and hole states} gives an essential contribution to $W(\varepsilon ,\varepsilon ')$ 
due to the matrix structure of the spectral density functions. After the integrations over $\bf p$-plane, one transforms (11) into 
\begin{eqnarray}
W(\varepsilon ,\varepsilon ')\equiv\Theta_{GR}w(\varepsilon /T,
-\varepsilon '/T) /T^2 , \\
\Theta_{GR}=\frac{\upsilon_{ac}s}{\upsilon^2}\frac{T}{\hbar}\left(\frac{T}{\pi\hbar\upsilon}\right)^2 , ~~~~~~ \upsilon_{ac}=\frac{D^2T}
{4\hbar^2\rho_s\upsilon s^2} ,  \nonumber
\end{eqnarray}
where we separated the dimensionless kernel, $w(\xi ,\xi ')$, and the 
factor, $\Theta_{GR}$, which is written for the case of the deformation interaction of carriers with the in-plane acoustic modes, see Refs. 10 and 15. Here $D$ is the deformation potential, $s$ is the sound velocity, 
and $\rho_s$ is the sheet density of graphene. At room temperature and typical other parameters \cite{9} we obtain
$\upsilon_{ac}\simeq 0.96\times 10^6$ cm/s and $\Theta_{GR}\simeq 5.06\times 10^{19}$
cm$^{-2}$s$^{-1}$ (notice, that $\upsilon_{ac}\propto D^2$ and we used $D\simeq$ 12 eV).
The dimensionless kernel is plotted in Fig. 3 and the probability $W(\varepsilon ,\varepsilon ')$ is suppressed fast if $(\varepsilon -\varepsilon ')/T\geq$0.15. This cut-off factor is determined by the weak 
ratio $s/\upsilon\simeq$1/137 mainly while parameter $g$ determines a peak value
of $W(\varepsilon ,\varepsilon ')$, c.f. Figs. 3a and 3b. 

The generation-recombination rate (6) is written through (11) and (12) with
the use of the dimensionless variables $\xi =\varepsilon /T$ and $\xi ' =\varepsilon '/T$:
\begin{eqnarray}
\left(\frac{dn}{dt}\right)_{ac}=\Theta_{GR}\int\limits_0^\infty d\xi\int
\limits_0^\infty d\xi 'w(\xi ,\xi ') \\ 
\times\frac{e^{-\xi '-\mu_t^> /T}-e^{-\xi '-\mu_t^< /T}}{\left[\exp\left(
\xi -\frac{\mu_t^>}{T}\right) +1\right] \left[\exp\left(\xi '-\frac{\mu_t^<}
{T}\right) +1\right]}  \nonumber 
\end{eqnarray}
For typical concentrations of carriers, $\mu_t^<$ and $\mu_t^>$ exceed 
0.15$T$ and one can simplify the rate as follows:
\begin{eqnarray}
\left(\frac{dn}{dt}\right)_{ac}\approx\Theta_{GR}\frac{w_g\left[ e^{-
\mu_t^> /T}-e^{-\mu_t^< /T}\right]}{\left[e^{-\mu_t^> /T}+1\right] 
\left[e^{-\mu_t^</T} +1\right]} , \nonumber \\
w_g=\int\limits_0^\infty d\xi\int\limits_0^\infty d\xi 'w(\xi ,\xi ')
 ~~~~~~~
\end{eqnarray}
where the averaged over energies kernel $w_g$ is plotted versus $g$ in 
Fig. 4, together with a simple parabolic fit. For the disorder level corresponding to the resistance $\sim$5 k$\Omega$ per square, one obtains $w_g\simeq$0.02.
\begin{figure}
\begin{center}
\includegraphics{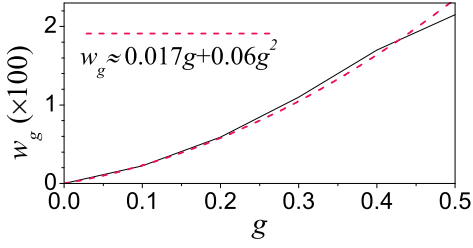}
\end{center}
\addvspace{-0.5 cm}
\caption{(Color online) Averaged kernel $w_g$ versus coupling 
constant $g$. }
\end{figure}

\section{Results}
In this section we analyze the concentration balance equation $dn_t/dt
=(dn/dt)_{ac}$, where the right-hand side of Eq. (14) is written through 
$\psi_{t}^> =\mu_t^>/T$ and $\psi_{t}^< =\mu_t^>/T$, together with the 
initial conditions $\psi_{t=0}^> =\psi_0^>$ and $\psi_{t=0}^< =\psi_0^<$
determined through the initial concentrations $n_{t=0}$ and 
$\overline{n}_{t=0}$ according to Eq. (4). Variables $\psi_{t}^>$ and 
$\psi_{t}^<$ are connected through the electroneutrality condition
\begin{equation}
\int\limits_0^\infty d\varepsilon \rho_\varepsilon\left(\frac{1}
{e^{\varepsilon /T-\psi_t^>}+1}-\frac{1}{e^{\varepsilon /T+\psi_t^<}
+1} \right) = n_s 
\end{equation}
and below we consider the cases of an intrinsic graphene ($n_s=0$) and a $n$-type heavily-doped graphene ($n_s>0$).

\subsection{Intrinsic graphene}
For the case under consideration, $\psi_{t}^> =-\psi_{t}^<\equiv\psi_{t}$
and the concentration of electrons (or holes, because now $n_{t}=
\overline{n}_{t}$) is given by $n_t =\int_0^\infty d\varepsilon \rho_\varepsilon\left[\exp (\varepsilon /T -\psi_t )+1\right]^{-1}$, 
so that $n_t$ and $\psi_t$ are connected through
\begin{equation}
\frac{dn_t}{dt}=\frac{d\psi_t}{dt}\int\limits_0^\infty\frac{d\varepsilon 
\rho_\varepsilon}{1+\cosh\left(\varepsilon /T-\psi_t\right) } .
\end{equation}
As a result, the concentration balance equation is derived from Eq. (13)
as
\begin{equation}
\frac{dn_t}{dt}=-\Theta_{GR}w_g\tanh\left(\frac{\psi_t}{2}\right)
\end{equation}
and Eqs. (16) and (17) are transformed into the first-order differential equation for $\psi_{t}$ with the initial condition $\psi_{t=0}=\psi_0$ 
where $\psi_0$ is determined through the $n_{t=0}$. The implicit solution of this equation takes form:
\begin{eqnarray}
\nu_{GR}t=\int\limits_{\psi_t}^{\psi_0}d\psi F(\psi ),  ~~~~ \nu_{GR}
=w_g\frac{\upsilon_{ac}s}{\pi\upsilon^2}\frac{T}{\hbar} , \\
F(\psi )=\tanh\left(\frac{\psi}{2}\right)\int\limits_0^\infty
\frac{d\xi r_\xi}{1+\cosh (\xi -\psi )} . \nonumber 
\end{eqnarray}
Here $r_\xi =\rho_{\xi T}/\overline\rho_T$ is the dimensionless density of states and the temporal evolution of $n_t$ is described through the characteristic rate $\nu_{GR}$ and the dimensionless function $F(\psi )$. At room temperature and at $w_g\simeq$0.02, one obtains $\nu_{GR}\simeq 1.85\times 10^7$ s$^{-1}$ for the parameters used.

Figure 5 shows the transient evolution of $n_t$ normalized to the equilibrium
concentration $n_{eq}$, for the cases of recombination or generation of
carriers, if $n_{t=0}>n_{eq}$ or $n_{t=0}<n_{eq}$, respectively. The relaxation becomes suppressed if the disorder level decreases both due to
a slowness of dependency on $\nu_{GR}t$, c.f. curves for $g=$0.5 and 0.25
in Fig. 5, and, mainly, due to the relation $\nu_{GR}\propto w_g$, see 
Fig. 4. Within a 5\% accuracy, the evolution of $n_t$ can be fitted by the exponential dependencies
\begin{equation}
n_t\approx n_0 +n_T\left[ 1-\exp (-\alpha\nu_{GR} t)\right]
\end{equation}
with the parameter $\alpha$ varying between 0.16 and 0.28 depending on the
initial conditions. The corresponding times, $(\alpha\nu_{GR})^{-1}$, vary
between 310 and 180 ns for $g\simeq$0.5. This time scale is comparable to 
the radiative recombination times, see Appendix.
\begin{figure}
\begin{center}
\includegraphics{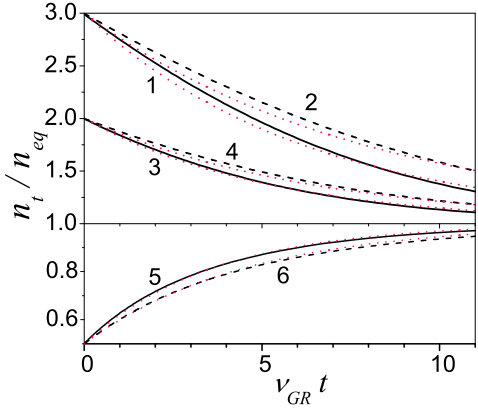}
\end{center}
\addvspace{-0.5 cm}
\caption{(Color online) Transient evolution of concentration $n_t$ at
different initial conditions: $n_{t=0}=3n_{eq}$ (1,2), $n_{t=0}=2n_{eq}$,
(3,4) and $n_{t=0}=0.5n_{eq}$ (5,6) for coupling parameters $g=$0.25 
(1, 3, 5) and $g=$0.5 (2, 4, 6). Dotted curves correspond to exponential fits (19), with $\alpha =$0.16 (1), 0.125 (2), 0.19 (3), 
0.155 (4), 0.22 (5), and 0.275 (6). }
\end{figure}

\subsection{Heavily-doped graphene}
We turn now to the case of a heavily doped graphene, when $\psi_t^>\gg 1$
and it is convenient to introduce a weak variation $\delta\psi_t =\psi_t^> 
-\psi_s$ where $\psi_s$ corresponds to the equilibrium case. Neglecting 
a hole concentration and using the step function in $c$-band, one obtains $n_s$ from Eq. (15):  
\begin{equation}
n_s\approx\int\limits_0^{\psi_s T}d\varepsilon \rho_\varepsilon .
\end{equation}
As a result, $\delta\psi_t$ and $\psi_t^<$ are connected by 
the electroneutrality condition (15) as follows
\begin{equation}
\delta\psi_t\approx\frac{\overline\rho_T}{\rho_{\varepsilon =\psi_s T}}\int\limits_0^\infty\frac{d\xi r_\xi}{1+\exp\left(\xi +\psi_t^<\right)} ,
\end{equation}
where the ratio $\overline\rho_T /\rho_{\varepsilon =\psi_s T}$ can
be found from Fig. 2a. Using Eq. (14) we transform the concentration 
balance equation into the form:
\begin{equation}
\frac{d\delta\psi_t}{dt}=-\frac{\nu_{GR}\overline\rho_T}{2
\rho_{\varepsilon =\psi_s T}[1+\exp\left(\psi_t^< \right) ]} . 
\end{equation} 
Substituting the relation (21) into Eq. (22) one obtains the first-order differential equation for $\psi_t^<$, with the implicit solution
\begin{equation}
\nu_{GR}t =\int\limits_{\overline\psi_0}^{\psi_t^<}d\psi\int\limits_0
^\infty\frac{d\xi r_\xi\left( 1 +e^{\psi}\right)}{ 1+\cosh 
\left( \xi +\psi\right) } ,
\end{equation}
where $\overline\psi_0=\psi_{t=0}^<$ appears from the initial condition.
Notice, that the factor $\overline\rho_T /\rho_{\varepsilon =\psi_s T}$
drops out from the solution (23), i. e. the transient process under consideration does not depend on the doping level because the only low-energy states are involved in the interband transitions.
\begin{figure}
\begin{center}
\includegraphics{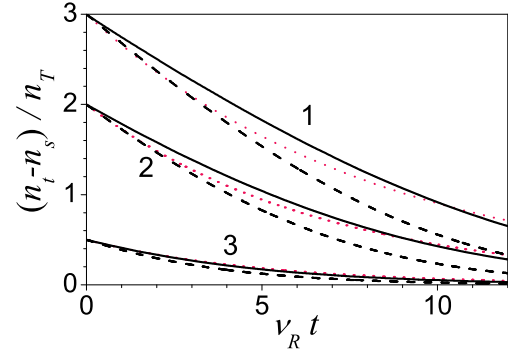}
\end{center}
\addvspace{-0.5 cm}
\caption{(Color online) Transient evolution of concentration $n_t -n_s$ 
at different initial conditions: $n_{t=0}-n_s=3n_T$ (1), $n_{t=0}-n_s=
2n_T$, (2) and $n_{t=0}-n_s=0.5n_T$ (3). Solid and dashed curves correspond to coupling parameters $g=$0.5 and $g=$0.25, respectively. Dotted curves  correspond to the exponential fits. }
\end{figure}

Further, we plot the transient evolution of the hole concentration
$\overline{n}_t=n_t -n_s$ determined through $\psi_t^<$ according to 
$\overline{n}_t=\int_0^\infty d\varepsilon \rho_\varepsilon\left[
\exp (\varepsilon /T +\psi_t^< )+1\right]^{-1}$. Figure 6 shows the 
concentration $\overline{n}_t$ versus dimensionless time, $\nu_{GR}t$, 
for the initial conditions written through $n_T$. Similarly to the 
undoped case, the exponential fits $(n_t -n_s) /(n_{t=0}-n_s)\approx
\exp (-\beta\nu_{GR}t)$ with $\beta\simeq$0.12 (1), 0.15 (2), and 0.2 
(3) describe the transient evolution with an accuracy $\sim$10\% if 
$\nu_{GR}t<$10. An enhancement of recombination takes place at tails of 
transient evolution, if $\nu_{GR}t>$10. Since the relaxation rate
increases with the disorder level, $\nu_{GR}\propto w_g$, the recombination process becomes faster in spite of an opposite dependency on $\nu_{GR}t$ 
in Fig. 6. The relaxation times, $\sim (\beta\nu_{GR})^{-1}$, vary
between 410 and 240 ns for $g=$0.5 and different initial conditions. Once
again, the recombination scale is comparable to the radiative recombination process shown in Fig. 7b, Appendix.

\section{SUMMARY AND CONCLUSIONS}
We have examined the new channel for interband generation-recombination process of
carriers in a disordered graphene via acoustic phonons. The efficiency of transitions increases with the  disorder level and concentration of nonequilibrium carriers as well as with temperature. We have found that the relaxation rate belongs to submicrosecond range for the samples with typical disorder level at room temperature. 

Let us discuss the assumptions used in the presented calculations. The
main restriction of the results is the description of the response in 
the framework of the quasiequilibrium approach, with different chemical potentials in $c$- and $v$-bands but the same temperature due to the fast energy relaxation caused by phonon and carrier-carrier scattering processes.
We also restrict ourselves by the simplest model of the short-range disorder. 
By analogy with the description of transport phenomena, \cite{1,9,11} more
complicated calculations for finite-range disorder should give similar results. But the case of impurities with a low-energy resonant level, which 
was discussed recently in Refs. 16, requires a special consideration. 
We considered the deformation interaction of carriers with longitudinal acoustic modes \cite{9,14} neglecting scattering by surface phonons of 
the substrate in agreement with the experimental data. \cite{16} Such a
contribution can only restrict the energies under consideration
because of the lower surface phonon energy ($\sim$55 meV for the SiO$_2$ 
substrate). Since the non-diagonal components give a weak contribution to the concentration balance equation under consideration, \cite{10} we take into account only diagonal components of the density matrix $f_{\alpha t}$ while evaluating of the generation-recombination rate. The simplifications mentioned above do not change either the peculiarities of the generation-recombination processes or the numerical estimates of relaxation times given in Sec. III.

Next, we briefly consider some possibilities for experimental verification 
of the mechanism of interband transitions suggested. It is clear from a
comparison of the results in Sec. III and in Appendix that interband transitions via acoustic phonons and via thermal radiation can be separated 
due to different temperature and concentration dependencies of damping.
A possible contribution of the disorder-induced Auger process is beyond of
our consideration and requires a special study. In contrast to the ultrafast
optical measurements applied for the study of the relaxation and recombination 
of high-energy carriers, \cite{2} a transient evolution of concentration 
over time scales $\sim$100 ns can be measured directly (e.g. in Ref. 5 the 
transient response under abrupt switching on of a dc field lasts up to 
hundreds of nanoseconds). But under a verification of the slow process examined, 
a possible contact injection or a trapping into substrate states should 
be analyzed.

To conclude, we believe that the generation-recombination via 
acoustic phonons can be verified experimentally and more detailed numerical calculations are necessary in order to separate this mechanism from other contributions. The results obtained will stimulate a further study of the
generation-recombination processes which are essential in many transport 
and optical phenomena far from equilibrium.

\section*{ACKNOWLEDGMENT}
This paper is based upon work supported by the National Science
Foundation under Grant No DMR 0907126.

\appendix* 
\section{Radiative transitions}
Below we describe the generation-recombination processes which are 
associated with the interband transitions induced by the thermal 
radiation and evaluate the radiative relaxation rate for the weak
disorder case, $g\ll 1$. The corresponding collision integral was 
evaluated in Ref. 9 and the kinetic equation for the electron 
distribution $f_{ept}$ takes the form 
\begin{equation}
\frac{\partial f_{ept}}{\partial t} =\nu_p^{(R)}\left[ N_{2\upsilon p/T}
\left( 1-f_{ept}-f_{hpt} \right) -f_{ept}f_{hpt}\right] ,
\end{equation}
where $N_{2\upsilon p/T}$ describes the Planck distribution of the 
thermal photons at temperature $T$. The hole distribution can be obtained 
from the condition $\partial (f_{ept}+f_{hpt})/\partial t=0$. The 
interband absorption or emission of photons are described by the first 
or second terms in the right-hand
side of Eq. (A.1) and are responsible for the generation or recombination processes. The rate of spontaneous radiative transitions is given by 
$\nu_p^{(R)}=\upsilon_R p/\hbar$ where we have introduced the 
characteristic velocity $\upsilon_R\simeq$41.6 cm/s for graphene surrounded 
by SiO$_2$ layers. Similar to Eq. (6) contribution of the radiative 
collision integral from (A.1) into the concentration balance equation 
takes the form $(dn/dt)_{ac}=(4/L^2 )\sum\nolimits_{\bf p}\nu_p^{(R)}\left[ 
N_{2\upsilon p/T}\left( 1-f_{ept}-f_{hpt} \right) -f_{ept}f_{hpt}\right] $.

For the case of an intrinsic graphene, the balance equation is written 
by analogy with Sect. III A through $\psi_t =\mu_t /T$:
\begin{equation}
\frac{d\psi_t}{dt} =-\nu_R \frac{F_i(\psi_t )}{N(\psi_t )}, ~~~~ 
\nu_R =\frac{2\upsilon_R}{\upsilon}\frac{T}{\hbar}
\end{equation}
where $\nu_R^{-1}\approx$30 ns is the radiative recombination time 
at room temperature. The functions $F_i(\psi )$ and $N(\psi )$ are 
given by
\begin{eqnarray}
F_i(\psi )=\int\limits_0^\infty\frac{d\xi\xi\left( 1-e^{-2\psi}\right)}{\left( 1-e^{-2\xi} \right)\left( e^{\xi -\psi}+1\right)^2}  \nonumber \\
N(\psi )=\int\limits_0^\infty\frac{d\xi\xi}{1+\cosh (\xi -\psi )} 
\end{eqnarray}
and the implicit solution of Eqs. (A.2), (A.3) is given by the similar 
to Eq. (18) formula: 
\begin{equation}
\nu_R t=\int_{\psi_0}^{\psi_t}d\psi\frac{N(\psi )}{F_i(\psi )} .
\end{equation} 
In Fig. 7a we plot the transient evolution of concentration versus the
dimensionless time, $\nu_R t$ for the same initial conditions as in Fig. 5.
These transient dependencies are described by the exponential decay
given by Eq. (19) with $\alpha\approx$0.25 for all cases. Thus, one obtains
the radiative recombination time $\sim 4/\nu_R\approx$120 ns which does not depend on an initial concentration.  
\begin{figure}
\begin{center}
\includegraphics{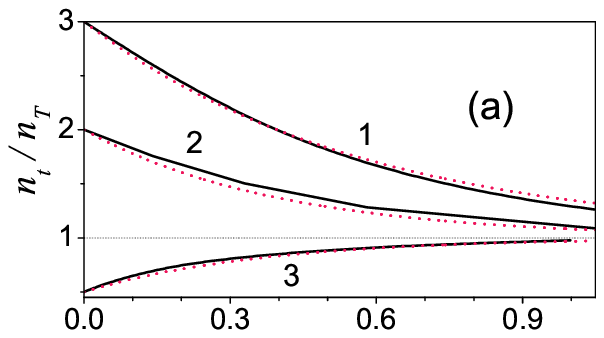}
\includegraphics{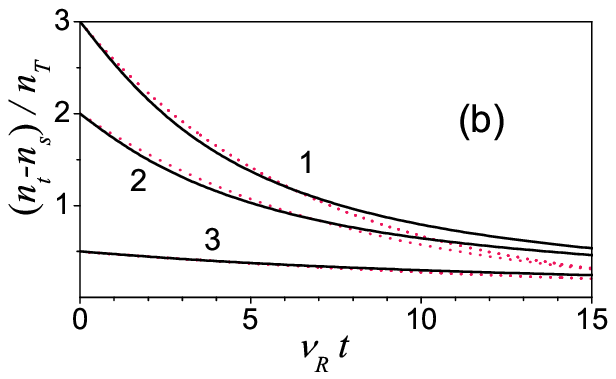}
\end{center}
\addvspace{-0.5 cm}
\caption{(Color online) Transient evolution of concentration due to 
interband radiative transitions for intrinsic (a) and heavily-doped (b)
graphene at different initial conditions: $n_{t=0}=3n_T$ (1), 
$n_{t=0}=2n_T$ (2) and $n_{t=0}=0.5n_T$ (3). Dotted curves 
correspond to exponential fits. }
\end{figure}

For the case of doped graphene, the concentration balance equation (22)
should be replaced by
\begin{eqnarray}
\frac{d\delta\psi_t}{dt}=-\frac{\nu_R}{2}F_d(\psi_t^< ), \\
F_d(\psi )=\int\limits_0^\infty\frac{d\xi\xi^2}{\left( 1-e^{-2\xi}\right)\left( e^{\xi +\psi}+1\right)}  \nonumber
\end{eqnarray} 
while the relation between $\delta\psi_t$ and $\psi_t^<$ takes form 
[c.f. Eq. (21)]
\begin{equation}
\delta\psi_t\approx\frac{1}{\psi_s}\int\limits_0^\infty\frac{d\xi\xi}{1+\exp\left(\xi +\psi_t^<\right)} .
\end{equation}
As a result, the equation for $\psi_t^<$ has the only difference from 
Eq. (2) due to the replacement $F_i(\psi )$ by $F_d(\psi )$. The implicit
solution of Eqs. (A.5) and (A.6) is given by (A.4) with the same replacement. In Fig. 7b we plot the transient evolution of hole concentration, 
$\overline{n}_t/n_T$, for the same initial conditions as in Fig. 6. The corresponding exponential fits are determined by the coefficients $\beta\simeq$0.15 (1), 0.125 (2), and 0.06 (3), i.e. the relaxation rate 
depends on hole concentration. At room temperature the radiative recombination time $(\beta\nu_R )^{-1}$ corresponds to the time interval between 190 and 480 ns.


\begin{thebibliography}{99}
\bibitem{1} 
N. M. R. Peres, Rev. Mod. Phys. {\bf 82}, 2673 (2010).

\bibitem{2} 
M. Orlita and M. Potemski, Semicond. Sci. Technol. {\bf 25} 063001 (2010);
F. Bonaccorso, Z. Sun, T. Hasan, and A. C. Ferrari, Nature Photonics 
{\bf 4}, 611 (2010).

\bibitem{2a}
S. Rumyantsev, G.  Liu,W.  Stillman, M. Shur, and A. A. Balandin, J. Phys.: Condens. Matter {\bf 22},  395302 (2010). 

\bibitem{3} 
A. V. Germanenko and G. M. Minkov, Phys. Stat. Sol. (b) {\bf 184},
9 (1994).

\bibitem{4} 
F. Rana, P. A. George, J. H. Strait, J. Dawlaty, S. Shivaraman,
Mvs Chandrashekhar, and M. G. Spencer, Phys. Rev. B {\bf 79}, 115447
(2009); F. T. Vasko, {\it ibid.} {\bf 82}, 245422 (2010).


\bibitem{5} 
P. N. Romanets and F. T. Vasko, Phys. Rev. B {\bf 83}, 205427 (2011).

\bibitem{6} 
M. S. Foster and I. L. Aleiner, Phys. Rev. B {\bf 79}, 085415  (2009);
D. M. Basko, S. Piscanec, and A. C. Ferrari, Phys. Rev. B {\bf 80}, 165413 
(2009).

\bibitem{7} 
F. Rana, J. H. Strait, H. Wang, and C. Manolatou, arXiv:1009.2626; 
T. Winzer, A. Knorr, and E. Malic, Nano Letters {\bf 10}, 4839 (2010);
F. Rana, Phys. Rev. B {\bf 76}, 155431 (2007).

\bibitem{8} 
F. T. Vasko and V. Ryzhii, Phys. Rev. B {\bf 77}, 195433 (2008).

\bibitem{9} 
F. T. Vasko and V. Ryzhii, Phys. Rev. B {\bf 76}, 233404 (2007).

\bibitem{10} 
F. T. Vasko and O. E. Raichev, \emph{Quantum Kinetic Theory and
Applications} (Springer, New York, 2005).

\bibitem{11} 
T. Ando, J. Phys. Soc. Jpn. {\bf 75}, 074716 (2006);
P. M. Ostrovsky, I. V. Gornyi, and A. D. Mirlin, Phys. Rev. B
{\bf 74}, 235443(2006).

\bibitem{12} 
T. Stauber, N. M. R. Peres, and A. H. Castro Neto, Phys. Rev. B {\bf 78}, 085418 (2008).

\bibitem{13} G. D. Mahan, \emph{Many-Particle Physics} (Plenum Press, N.Y.,
1990).

\bibitem{14} 
N. M. R. Peres, J. M. B. Lopes dos Santos, and T. Stauber, Phys. 
Rev. B {\bf 76}, 073412 (2007).

\bibitem{15} 
V. M. Pereira, J. M. B. Lopes dos Santos, and A. H. Castro Neto, Phys.
Rev. B {\bf 77}, 115109 (2008); B. Dora, K. Ziegler, and P. Thalmeier, ibid.
{\bf 77}, 115422 (2008); 
M. Titov, P. M. Ostrovsky, I. V. Gornyi, A. Schuessler, A. D. Mirlin,
Phys. Rev. Lett. {\bf 104}, 076802 (2010). 

\bibitem{16}
A. Barreiro, M. Lazzeri, J. Moser, F. Mauri, and A. Bachtold, Phys.
Rev. Lett. {\bf 103}, 076601 (2009); S. Fratini and F. Guinea, Phys. Rev. 
B {\bf 77}, 195415 (2008).

\end{thebibliography}
\end{document}